\documentclass[submission, Phys]{SciPost}
\usepackage{wrapfig}
\usepackage{mathrsfs}

\usepackage{amssymb}

  \let\g=\gamma

\let\s=\sigma

  \let\th=\theta

\newcommand{\C}{\mathcal{S}}
\newcommand{\Ov}{\mathcal{T}}

\def\to{\rightarrow}

\newcommand{\beq}{\begin{equation}} \newcommand{\eeq}{\end{equation}}

\begin{document}

\begin{center}{\Large \textbf{
Critical energy landscape of linear soft spheres
}}\end{center}

% TODO: write the author list here. Use initials + surname format.
% Separate subsequent authors by a comma, omit comma at the end of the list.
% Mark the corresponding author with a superscript *.
\begin{center}
Silvio Franz\textsuperscript{1},
Antonio Sclocchi\textsuperscript{1*},
Pierfrancesco Urbani\textsuperscript{2}
\end{center}

\begin{center}
{\bf 1} Universit\'e Paris-Saclay, CNRS, LPTMS, 91405, Orsay, France
\\
{\bf 2} Universit\'e Paris-Saclay, CNRS, CEA, Institut de physique th\'eorique, 91191, Gif-sur-Yvette, France.
\\
*antonio.sclocchi@universite-paris-saclay.fr
\end{center}

%\begin{center}
%\today
%\end{center}

% For convenience during refereeing: line numbers
%\linenumbers

\section*{Abstract}
{\bf
We show that soft spheres
interacting with a linear ramp potential when overcompressed beyond
the jamming point fall in an amorphous solid phase which 
is critical, mechanically marginally stable and share many 
features with the jamming point itself. 
In the whole phase, the relevant local minima of the potential energy landscape 
display an isostatic contact network of perfectly touching spheres 
whose statistics is controlled by an infinite lengthscale.
Excitations around such energy minima are non-linear,
system spanning, and characterized by a set of non-trivial critical
exponents. We perform numerical simulations to measure their values
and show that, while they coincide, 
within numerical precision, with the critical exponents appearing at jamming, 
the nature of the corresponding excitations is richer. 
Therefore, linear soft spheres appear as a novel class of finite dimensional systems
that self-organize into new, critical, marginally stable, states.
}

\vspace{10pt}
\noindent\rule{\textwidth}{1pt}
\tableofcontents\thispagestyle{fancy}
\noindent\rule{\textwidth}{1pt}
\vspace{10pt}

\section{Introduction}

Since more than twenty years, the ideal jamming
points of systems of frictionless spheres  have shaped our thinking of
low temperature glasses, suggested principles underlying amorphous
rigidity, and provided mechanisms to rationalize low energy
excitations in glasses \cite{LN98, LN10}.  Topic feature of 
packings at jamming is mechanical marginal stability. 
The number of contacts between the spheres is isostatic, in $d$ dimensions 
each sphere has on average 2$d$ contacts, that is the least one for which the system can sustain
pressure \cite{OLLN02, OSLN03,CWBC98, TW99}. 
As such, mechanical marginal stability brings about criticality and
diverging lengthscales \cite{LNSW10}. 
The jamming point is a critical point characterized by a set of critical exponents describing both the behavior of bulk physical quantities, such as pressure, energy and contacts \cite{OSLN03, GLS16} as well as the microstructure of amorphous packings \cite{LDW13, CCPZ12, CCPZ15}. 
In particular, a common characterization is
provided by local statistics of contact forces and interparticle
distances. Marginal stability implies power law behavior of
the distribution of these quantities at small argument \cite{LDW13, CCPZ12} and predicts a
non trivial relation between the corresponding exponents \cite{Wy12}. These
exponents have been computed exactly in \cite{CKPUZ14NatComm, CKPUZ14JSTAT} 
and have been shown to agree -within numerical precision- with numerical simulations 
of hard and soft spheres in various  physical dimensions
\cite{CKPUZ17}.
The universality class of marginally stable jamming points has been
further shown to go beyond finite dimensional sphere systems and to
encompass more generally a large class of continuous constraint
satisfaction problems in machine learning and computer science
\cite{FP16, FPUZ15,FPSUZ17,FHU19}. 
For soft constraints, jamming points are isolated critical points: in general, typical
soft sphere systems (Harmonic or Hertzian spheres) compressed beyond the jamming point loose most of the
salient critical features of jamming, becoming
mechanically stable with a finite correlation length. 
In this paper, we show that if we fine tune the soft sphere interaction
potential - choosing it as a linear ramp - we can get a new amorphous solid phase which is mechanically
marginally stable and critical for all densities beyond the jamming
point.  Furthermore, the emerging marginal stability is richer that the one appearing at the boundary jamming transition, with additional system spanning non-linear excitations.

\section{Model and main results}
We consider a set of $N$ spheres in $d$ dimensions whose centers are $d$-dimensional vectors denoted by $\{{\bf x}_i\}_{i=1, \ldots, N}$.
We define a gap between two spheres, say $i$ and $j$, as $h_{ij} =
r_{ij}-\s_{ij}$, where we have denoted by $\s_{ij}$ 
the sum of the radii of the corresponding spheres and by $r_{ij}=|{\bf x}_{i} - {\bf x}_j|$ the distance between their centers. 
In the overcompressed phase, above jamming, spheres cannot be arranged without creating overlap between them. Therefore one typically defines a pure
power interaction potential $v_\alpha(h_{ij})=f_c (h_{ij})_+^\alpha$
where $x_+=|x|\theta(-x)$. $f_c$ is a constant that essentially sets the unit of forces. 
Common choices for the penalty exponent $\alpha$ are $\alpha=2$ or $\alpha=5/2$,
corresponding respectively to Harmonic and Hertzian spheres.
If $\alpha>1$ the interaction potential is convex and differentiable at
$r_{ij}=\sigma_{ij}$, i.e. when spheres just touch. As a consequence, given a contact at jamming, an
infinitesimal normal force is enough to destabilize it and can cause an
overlap between the corresponding particles. Therefore, for $\alpha>1$ jamming is a
singular point in the phase diagram: as soon as the
spheres overlap, the system stabilizes. This implies that jamming criticality
is washed out when we enter in the overcompressed phase.
In this paper, we investigate the potential energy landscape (PEL) of soft spheres 
interacting through a linear ramp potential, obtained by setting
$\alpha=1$, above the jamming transition point. 
\begin{figure}
\centering
\includegraphics[width=0.5\columnwidth]{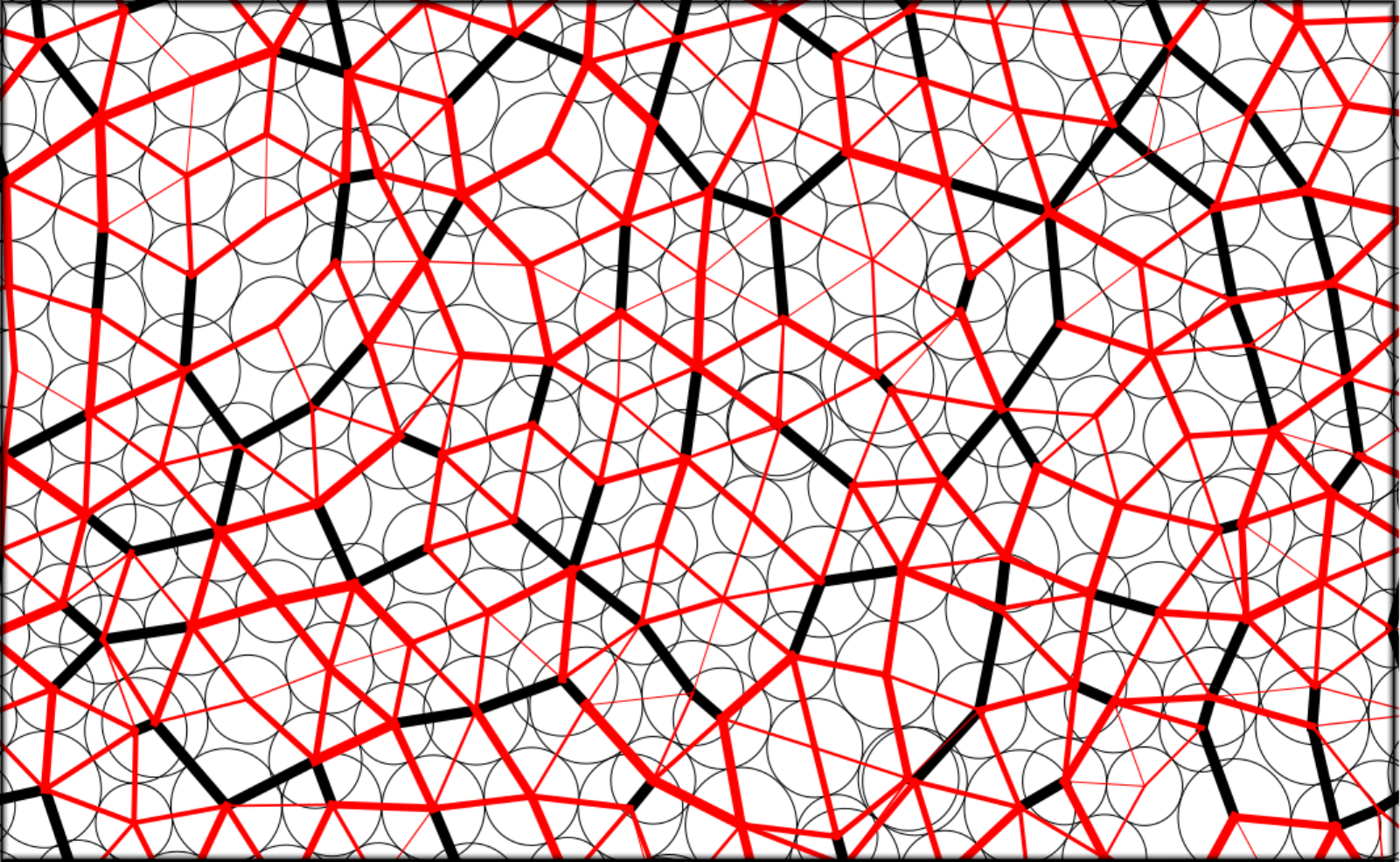}
\caption{A snapshot of a configuration of linear disks at packing fraction $\phi = 1$. The contact network is in red while the overlap network is in black. The thickness of the lines reflects the intensities of forces. While black lines carry all forces equal to one, red lines, associated to contacts, carry a varying force in the interval $[0,f_c=1]$.}
\label{fig:real_config}
\end{figure}
{We show that in this case the jammed phase presents new and unexpected features:
the linear ramp potential makes the overcompressed phase critical 
and marginally stable, characterized by a set of non-linear excitations whose nature is richer than the ones appearing at jamming.}
{The linear ramp} potential, which is at the boundary between convex and non-convex interparticle
potentials, presents important qualitative differences from the case
$\alpha>1$. First of all, it is non-differentiable: small forces applied to
contacts do not necessarily destabilize them. To induce an overlap, a total force greater than $f_c$ is necessary. In addition, the
modulus of the force generated by an overlap does not depend on the
extent of the overlap itself. 

We focus on systems of two and three dimensional polydisperse spheres and produce local minima by gradient descent minimization. 
Our main findings are:
\begin{itemize}
\item Accessible local minima of the PEL are
  isostatic. Even if there is a finite fraction of overlapping spheres making 
the total energy finite, there is also an isostatic number of pairs of spheres that just touch. 
We call \emph{interacting} spheres couples of spheres that either are in perfect contact (\emph{contacts}) $h_{ij}=0$, 
or that overlap (\emph{overlaps}) $h_{ij}<0$.
\item Contacts play a crucial role in the stability of the
  system. Their number is fixed to be exactly equal to the number of degrees of freedom and its fluctuations are suppressed, 
as it happens at jamming \cite{HUZ19}. We show that, as at jamming \cite{HLN17}, the spatial fluctuations of the local connectivity 
of the contact network are hyperuniform {implying that the variance of the number of contacts in a volume $V$ grows slower than $|V|$}. Conversely, the fluctuations of the number of overlaps  follow central limit theorem and spatial
  fluctuations of the overlap network are only uniform.
\item If we look at gap variables 
   and we focus on strictly positive and negative gaps,
we find that both distributions have a power law
  divergence for small argument (in absolute value). 
  The power law exponents controlling the divergence appear to be the same - within numerical precision - for both distributions  
and {very close} to the one of positive gaps at jamming.
\item Contacts can be associated with forces in the interval $[0,f_c]$. 
We measure the force empirical distribution and show that it displays two
singular pseudogaps, close to zero and close to $f_c$. The pseudogap
exponents appear to depend on the packing fraction close to jamming. However, if
we carefully separate the contribution of
``bucklers'', namely spheres that have $d+1$ interacting spheres \cite{CCPZ15}, from the bulk statistics, 
the pseudogaps are universal and
characterized by the same exponents in the whole jammed phase, far
from jamming. The values of the critical exponents appear to be the same -within
numerical precision - as the one of small force distribution at
jamming.
\end{itemize}
Isostaticity and critical behavior in the force and gap distributions have been shown to appear in the unsatisfiable phase of the spherical
perceptron optimization problem with linear cost function, which is a
mean field model for linear spheres \cite{FSU19}.  The main result of
the present work is that these properties appear to survive in a robust way when we go to finite dimension.
This implies that jammed packings of linear spheres are characterized
by diverging isostatic lengthscales and therefore are critical
even far from jamming in the compressed phase.
{Therefore they provide a new, richer example of self-organized critical, marginally stable, finite dimensional systems.}

\section{Numerical simulations}\label{sec:numerical_simulations}
The linear ramp $v_1(h)$ is a singular interparticle potential and therefore, both for the sake of theoretical comprehension and to perform numerical simulations, it is very useful to smooth the singularity out and to define a differentiable $\epsilon$-regularized potential between particles. From now on we set $f_c=1$\footnote{Note that $f_c$ sets only the overall scale of the maximal force and therefore we do not loose generality in setting it to one.}. We can define a regularized interparticle potential as
\begin{eqnarray}
  \label{eq:1}
  v_1(h;\epsilon)=\left\{
  \begin{array}{ll}
     0 & h>\frac \epsilon 2\\
     \frac{1}{2\epsilon}(h-\frac{\epsilon}{2})^2 & -\frac{\epsilon}{2}< h<\frac{\epsilon}{2}\\
     |h| & h<-\frac \epsilon 2
  \end{array}
  \right. 
\end{eqnarray}
which in the $\epsilon\to 0$ reduces to the linear ramp potential.
The potential energy of a system of $N$ spheres is therefore defined as $H_\epsilon (\underline x) = \sum_{i<j} v_1(h_{ij};\epsilon)$ and we want to study what happens in the $\epsilon \to 0$ limit.
In Eq.~(\ref{eq:1}), the non-differentiable point in the origin of $v_1(h)$ has been regularized by an arch of parabola of curvature $1/\epsilon$.  
Numerically, the regularization enables us to use gradient based routines. Theoretically, the model splits the degeneracy of forces at $h=0$ in an interval of order $\epsilon$, i.e. $-\epsilon/2<h<\epsilon/2$; it also allows to properly define the Hessian controlling minima's stability and to argue in favor of isostaticity.

\begin{figure}
\centering
\includegraphics[width=0.48\columnwidth]{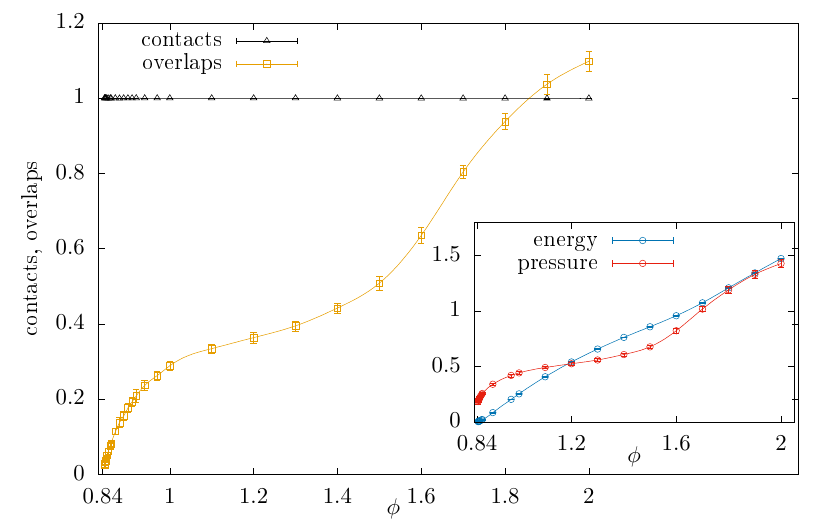}
\includegraphics[width=0.48\columnwidth]{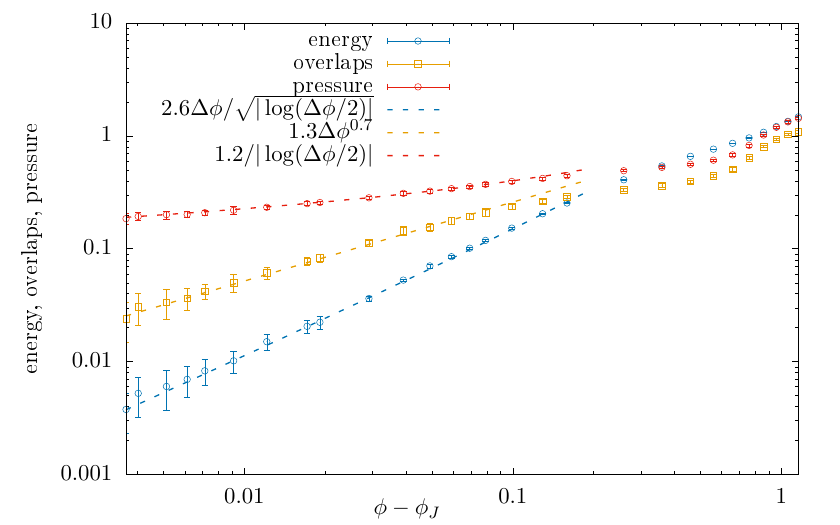}
\caption{\emph{Left Panel}. Main plot: the isostaticity index defined as $c=C/(Nd-d)$ and
  the fraction of overlaps defined as $n_O=O/(Nd-d)$ as a function of the
  packing fraction (we find the jamming transition at
  $\phi_J\simeq 0.84$). Inset:
  Behavior of energy and pressure for $\phi>\phi_J$. Energy, pressure and number of overlaps are increasing functions continuous at jamming.
  Data produced with system size $N=512$, dimensions $d=2$, averaged over $\sim 40$ samples for each point.
  \emph{Right Panel.} The behavior of pressure, energy and overlaps close to the unjamming transition. We attempted some logarithmic fits of the form $e\sim |\Delta \phi|/\sqrt{\log(\Delta\phi/2)}$, $p\sim 1/\sqrt{\log(\Delta\phi/2)} $ and $n_O\sim  |\Delta \phi|^{\nu_e}$. The unjamming packing fraction $\phi_J$ is extracted from the fit of the energy. }
\label{fig:iso}
\end{figure}

We consider systems of $N$ up to $4096$ disks in dimension $d=2$ and $N$ up to $1024$ spheres in dimension $d=3$, inside a $d-$dimensional box of side-length $L$ with periodic boundary conditions. 
The particles' radii $R_{\{i=1...N\}}$ are random uniformly distributed between the values $1-p$ and $1+p$, with polydispersity $p=0.2$.
The side-length $L$ of the box is set by the volume density $\phi = \sum_{i=1}^{N}k_d R_i^d / L^d$, with $k_d = \pi^{d/2}/\Gamma(1+d/2)$ where $\Gamma(x)$ is Euler gamma function.
Starting from a random configuration of particles' positions, we minimize the energy of the system $H_\epsilon (\underline x) = \sum_{i<j} v_1(h_{ij};\epsilon)$ with the regularized interparticle interaction potential  $v_1(h_{ij};\epsilon)$ defined in eq. (\ref{eq:1}).
The first minimization is run with $\epsilon = 10^{-2}$ using FIRE minimization algorithm \cite{fire-BK06}. 
From the obtained configuration, we reduce $\epsilon$ by a factor $2$ and repeat the minimization using an approximated conjugate-gradient method (we use the routine L-BFGS \cite{lbfgs-BN95}).
We repeat the procedure halving $\epsilon$ at each step and we stop at $\epsilon \sim 10^{-8}$. 
Using more accurate minimizers, it is possible to access to lower values of $\epsilon$.
We check that for the final value of $\epsilon$ we have that $\epsilon\ll \underset{ij}{\text{min}}|h_{ij}|$, where  $\underset{ij}{\text{min}}|h_{ij}|$ is the smallest non-zero gap of the configurations we are looking for.
With this procedure we meet the jamming transition
at packing fraction $\phi_J^{2d}\simeq 0.84$ and 
$\phi_J^{3d}\simeq 0.64$.
This procedure provides the set $\mathcal{C}$ of the $C=|\mathcal{C}|$ particles
pairs $\mu=\langle ij\rangle$, with $i<j$, that are in contact
(i.e. $-\epsilon/2<h_{ij}<\epsilon/2$) and the set $\mathcal{O}$ of the  $O=|\mathcal{O}|$ particles pairs $\mu=\langle ij\rangle$ that are overlapping (i.e. that have negative gaps $h_{ij}<-\epsilon/2$).
Associated to the contact pairs, there are the scalar contact forces $f_\mu=f_{ij}$ that form a $C-$dimensional vector $\vec{f}=\{f_{ij}\}$, while overlapping spheres exchange forces of intensity $1$, whose corresponding $O-$dimensional vector is simply $\vec{1}=[1,...,1]$.
The scalar contact forces $f_{ij}$ can be computed from the regularized potential of eq. \ref{eq:1} as
$f_{ij}= |h_{ij}-\frac{\epsilon}{2}|/\epsilon$, implying $0<f_{ij}<1$. 
Introducing the matrices $\C$ and $\Ov$, with dimensions $C \times Nd$ and $O\times Nd$ respectively, defined as $\C^{k\alpha}_{\langle ij\rangle} = (\delta_{jk}-\delta_{ik})n_{ij}^\alpha$, with $\langle ij\rangle\in \mathcal{C}$, and $\Ov^{k\alpha}_{\langle ij\rangle} = (\delta_{jk}-\delta_{ik})n_{ij}^\alpha$, with $\langle ij\rangle\in \mathcal{O}$, where $n_{ij}^\alpha$ is the $\alpha$-component of the versor $n_{ij}=(x_j-x_i)/|{\bf x}_j-{\bf x}_i|$,
we can compute the contact forces $f_{ij}$ also in an algebraic manner using the force-balance condition
\begin{equation}
\label{eq:fbc}
    \C^T \vec{f} = -\Ov^T \vec{1}.
\end{equation}
Notice that the system self-organizes in a way that the forces 
solving the linear system (\ref{eq:fbc}) lie in the interval $(0,1)$. 

In Fig.\ref{fig:real_config}, we show an example of a configuration we obtain through the numerical procedure just described.
In red and black we draw respectively the contact and overlap networks. 
In the following, we present data for $d=2$ and $N=512$ (unless otherwise specified). The data we got in $d=3$ are qualitatively similar to the $d=2$ case and therefore we report them in the appendix.

\subsection{The jammed phase}

In the jammed phase, for $\phi>\phi_J$, particles overlap and therefore 
the numbers of contacts $C$ and of overlaps $O$, the energy $E$ and the pressure $p$ are different from zero.  In two and in three dimensions, in all the minima we found, once removed the rattlers\footnote{Note that rattlers are present close to jamming but their density goes to zero very fast upon entering the jammed phase.},  $C$ is equal to the isostatic value $(N^*-1)d$, where $N^*$ is the number of spheres which are not rattlers.\footnote{In order to reach a minimum of the system, it is required a minimization algorithm with a spatial resolution of at least $\epsilon$, being $\epsilon$ the regularization parameter: meeting this requirement becomes more challenging when increasing the system size and reducing $\epsilon$.} On the other hand, $O$, $E$ and $p$ are continuous at jamming, having defined the pressure $p$ as $p= V^{-1} \sum_{i<j} |r_{ij}| f_{ij} /d$, where $f_{ij}$ is the force exchanged by the spheres $i$ and $j$ which can be $f_c$ in the case of spheres overlapping or it can be in the interval $(0,f_c=1)$ in the case of spheres in contact.

\begin{figure}
\centering
\includegraphics[width=0.5\columnwidth]{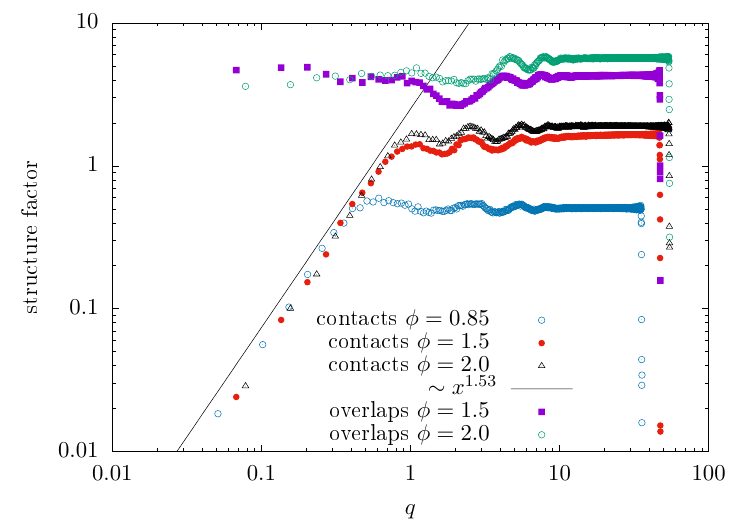}
\caption{Structure factor of the local connectivity of the network of contacts and overlaps. 
For small momentum, the structure factor of the contact network decreases down to zero
implying hyperuniformity in the fluctuations of connectivity. The exponent controlling the behavior of the structure factor appears to be close to $\sim 1.53$ which is the same found at jamming \cite{HLN17}. On the contrary, the connectivity of the overlap network is not hyperuniform. 
Data produced with system size $N=4096$, dimensions $d=2$, averaged over $44$ samples for $\phi=0.85$, over $50$ samples for $\phi=1.5$  and $48$ samples for $\phi=2$.}
\label{fig:structure}
\end{figure}

\begin{figure}
\centering
\includegraphics[width=0.45\columnwidth]{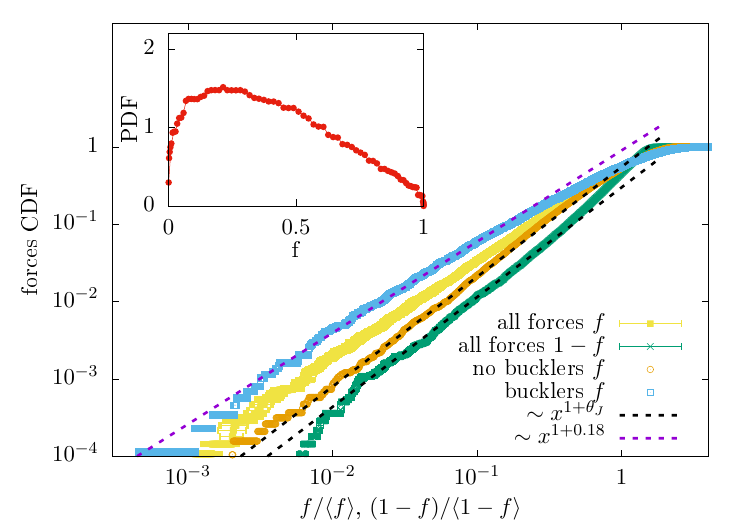}
\includegraphics[width=0.45\columnwidth]{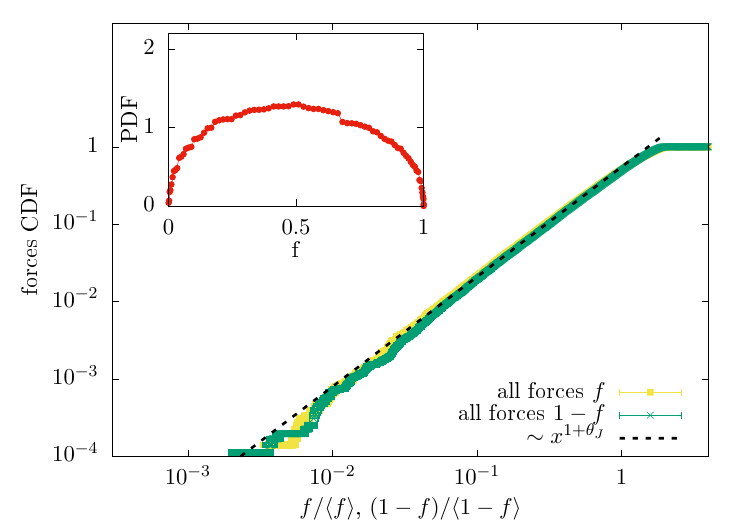}
\caption{Contact force distributions. \emph{Left panel}: the cumulative of the contact force distribution at $\phi=0.85$ in $2d$, close to the unjamming transition. We plot the cumulative both starting from the edge at $f=0$ and at $f=1$. While a blind statistics of forces is controlled by a hybrid power law exponent, once the effects of bucklers are removed we clearly observe power laws controlled by the mean field exponents, both close to $f=0^+$ and $f=1^-$. In the inset we plot the empirical probability distribution function. Data produced with system size $N=512$, dimensions $d=2$, averaged over $30$ samples. \emph{Right panel}: Cumulative distribution of contact forces close to zero and one at $\phi=2$ in $2d$, far from jamming. We observe that both distributions follow the mean field exponent. Our statistics is not sufficient to detect any localized excitations at this packing fraction and therefore in this case we consider directly all forces without separating  the contribution of bucklers from the analysis.
Data produced with system size $N=512$, dimensions $d=2$, averaged over $35$ samples.}
\label{fig:forces}
\end{figure}

This is shown in Fig.\ref{fig:iso}-left where we plot the isostaticity
index $c=C/[(N^*-1)d]$, for a $2d$ system. In the same figure, we plot
the overlap fraction $n_{\cal O}=O/[(N-1)d]$, and, in the inset, the energy
per particle and the pressure.  These quantities start from zero at
jamming and monotonically increase as the packing fraction grows. 
{We report in Fig. 2-right the behavior of energy, pressure and overlaps close to unjamming trying some
preliminary fits. Let us note that the scaling theory developed in \cite{OSLN03,Wyart,GLS16} becomes just marginal for the linear ramp potential and logarithmic behavior has to be expected. }
{In order to establish the precise form of the scalings close to unjamming one needs to consider proper decompression algorithms that allow to reduce sample to sample fluctuations close to the transition. This goes beyond the scope of this paper and will be the subject of a forthcoming work\cite{FSU20}.}
To characterize the networks of interaction, we study the fluctuations in the local contact number and overlap number.  Following \cite{HLN17}, we look at the 
local connectivity fluctuations of the networks of interactions by measuring the structure factors
\beq
S_{c,o}(q) = \frac 1N{\sum_{i,j=1}^N \langle \delta c_i \delta c_j e^{i q\cdot r_{ij}}}\rangle
\eeq
where $c_i$ represents the number of contacts or overlaps of particle $i$  for the contact or overlap structure factors respectively, $\delta c_i= c_i - \langle c\rangle$ is its local fluctuations and the angular brackets represent average over 
different minima.
We plot both structure factors in Fig.\ref{fig:structure} for different densities in $2d$. The behavior at small $q$
reveals a different behavior of fluctuations of contact and overlap numbers. 
The structure factor of the contact network decreases to zero at small argument, while the one of overlaps tends to a positive value. This signals that the fluctuations in contact number are hyperuniform in space, within a volume $V$, the square fluctuations of $C$ scale subextensively in $V$, while the ones of the overlap number are normal and scale as $V$. This difference is a manifestation of the  different role that contacts and overlaps play in the stability of the system. As the system is progressively compressed from the jamming point to higher densities, the networks self-organize keeping the 
number of contacts fixed while increasing the overlaps. 
As at regular jamming \cite{HLN17}, fluctuations of contact numbers away from isostaticity are suppressed and controlled by an infinite lengthscale.  
We note that the structure factor of the contact network shrinks to zero with a power law that is close to what is observed at jamming \cite{HLN17}.

We conclude by noting that while increasing the packing fraction, the fraction of overlaps displays {an inflection point around $\phi\sim 1.2$}. 
We empirically observe that at the same point the overlap network seems to undergo to a kind of percolation transition whose nature and properties are left for future investigations.

\subsection{Statistics of gaps}
\begin{figure}
\centering
\includegraphics[width=0.48\columnwidth]{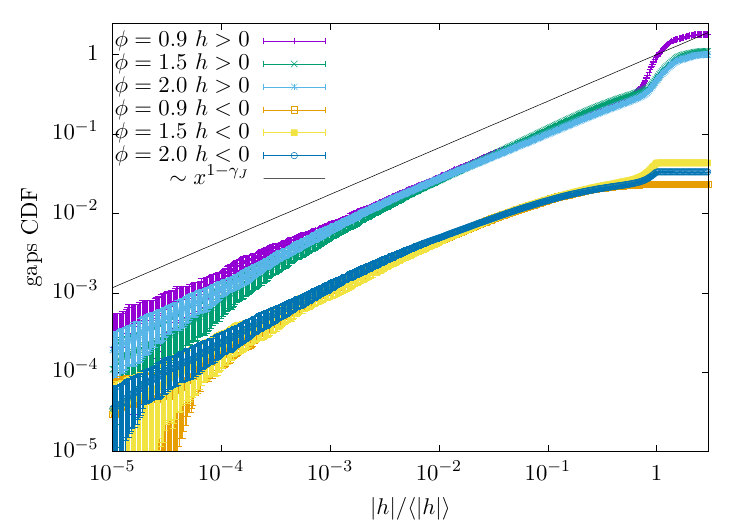}
\caption{The positive and negative cumulative gap distribution. The $y$-axis of the negative gap is rescaled by an artificial factor 0.1 to improve the readability of the figure.
We observe that both cumulative distribution appear to be described by the same power law exponent for small argument.
Data produced with system size $N=512$, dimensions $d=2$, averaged over $35$ samples for densities $\phi=0.9$ and $\phi=2.0$ and over $46$ samples for density $\phi=1.5$.}
\label{fig:gaps}
\end{figure}

Having established that the system is isostatic, 
it is natural to turn the attention to the distribution of non-zero gap variables, which at jamming provides an important characterization of criticality. 
While at jamming all gaps are positive or zero, here we also have 'negative gaps', quantifying the overlaps between particles. 
Both the distributions of positive and negative gaps appear to be singular at small argument.
In Fig. \ref{fig:gaps} we plot the cumulative distribution of both positive and negative gaps for several packing fractions beyond the jamming transition point. The small gap behavior  is controlled in both cases by a power law.
If we denote by $g_\pm(h)$ the positive and negative gap distribution, we have that
\begin{equation}
g_\pm(h)\sim |h|^{-\g_\pm}.
\end{equation}
at small argument. The two exponents coincide within the errors,  
$\g_+\approx \g_-$ and their numerical value appear to be independent of density and equal to the one of positive gaps at jamming $\g_\pm =\g_J\approx 0.41\ldots$  \cite{CKPUZ14NatComm}, as predicted by mean field theory \cite{FSU19}.

\subsection{Statistics of forces}
Local minima contain an isostatic number of contacts to which we can associate contact forces and study their empirical distribution. Scalar contact forces are naturally defined in the interval $[0, f_c = 1]$ and we observe that, as soon as we enter in the jammed phase, their distribution develops two pseudogaps close to the edges $f \sim 0, 1$ (see Fig. \ref{fig:forces}). We observe that the exponents controlling the two pseudogaps appear to depend smoothly on the density especially close to jamming and for $f\sim 0^+$. However, following \cite{CCPZ15} we perform a statistical analysis in which we remove bucklers, namely spheres interacting with $d + 1$ spheres. 
The result of the analysis is plotted in Fig. \ref{fig:forces} and we show that, independently from the packing fraction, the force distribution behaves as
\beq
p(f)\sim \left\{
  \begin{array}{ll}
f^{\th_-} & f\sim 0^+\\
(1-f)^{\th_+} & f\sim 1^-
\end{array}
\right.
\eeq
with $\theta_+\approx \theta_-\approx \theta_J$, where $\theta_J\simeq 0.42\ldots$ 
is the critical exponent controlling small forces between hard spheres at jamming \cite{CKPUZ14NatComm}.
In Fig. \ref{fig:forces}, we also plot the cumulative distribution function of bucklers' forces close to $f\sim 0$.
Again we see a power law behavior controlled -within numerical precision- by the same power law exponents controlling bucklers at jamming of hard spheres \cite{CCPZ15,LDW13}. Finally, we note that deep in the jammed phase, localized effects such as bucklers (but also rattlers) disappear (with the statistics we have access to) and we do not need to separate them from the statistics of forces to observe a critical power law with mean field exponent $\th_J$. 

\section{Non-linear marginal stability of linear soft spheres}
The exact solution of the perceptron optimization problem with a linear cost function \cite{FSU19}
provides a comprehensive mean-field framework that predicts the main features discussed in this paper, namely isostaticity, identity of exponents 
$\theta_+=\theta_-$, $\gamma_+=\gamma_-$, their numerical values and so on. This theory can be adapted straightforwardly to soft linear spheres in infinite dimension \cite{PUZ20}. In the next section we complement the mean field theory with marginal stability arguments. 

\subsection{Local stability}  The configurations of minima of the PEL at finite energy density contain overlapping particles.
It is easy to understand that there should also be pure contacts.  The forces corresponding to overlapping particles
are constant in modulus (equal to $f_c=1$) and, without contacts, only very symmetric configurations of particles would be mechanically stable. In fact, more generically, a number of contacts less than $d$ on a particle would require a highly symmetric configuration to be stabilized by only overlaps. The minimal number of contacts necessary to stabilize a single particle is therefore $d$, with a number of overlaps larger or equal to one (or with at least another contact). 
When we go from the jammed phase towards the jamming point, the number of overlaps vanishes and we recover that at jamming a number of contacts larger or equal to $d+1$ is required to block a sphere.  
Particles with $d+1$ interactions are prone to local excitations and are usually called bucklers.

\subsection{The regularized Hessian and isostaticity}
Local minima of the linear ramp potential are  anharmonic due to the singularity in the pairwise interaction potential.
However, one can consider the $\epsilon$-regularized potential and look at the Hessian of local minima in this case. This is indeed well defined and reads
{\medmuskip=0mu
\thinmuskip=0mu
\thickmuskip=0mu
\begin{equation}
  \label{eq:2}
 {\cal H}_{ij}^{ab}=
\begin{cases}
-\frac{1}{r_{ij}} v_1'(h_{ij};\epsilon)(\delta_{ab}-n_{ij}^a n_{ij}^b)-
  v_1''(h_{ij};\epsilon) n_{ij}^a n_{ij}^b & i\ne j\\
-\sum_{k\ne i} {\cal H}_{ik}^{ab} & i=j
\end{cases}
\end{equation}} 
\noindent with ${n}_{ij}^a=({x}_i^a-{x}_j^b)/|{\bf x}_i-{\bf x}_j|$,
$a,b=1,\ldots, d$.  Focusing on $i\neq j$, we
have the first term, often called prestress, which vanishes at
jamming, while we call the second term the elastic term.  Because of
the regularization, we have that
$v'_1(h;\epsilon)=(h-\frac{\epsilon}{2})/\epsilon\;{\bf I}[h\in [-\epsilon/2, \epsilon/2]]-{\bf 1}[h<
-\epsilon/2]$ and
$v''_1(h;\epsilon)={\bf I}[h\in [-\epsilon/2, \epsilon/2]]/\epsilon$, where we have
defined ${\bf I}[{\cal A}]$ the indicator function which is equal to
one if ${\cal A}$ is true and zero otherwise.  Notice that the Hessian
receives contributions both from overlaps and contacts. 
Overlaps
contribute just to the prestress. Contacts instead contribute both to
the prestress, with a finite term (notice that $(h-\frac{\epsilon}{2})/\epsilon$ is
actually the contact force), and to the elastic part with a term
proportional to $1/\epsilon$. This implies that for a variation of the
position of the particles such that $|\delta {x}_i|\lesssim \epsilon$,
the energy stored in the elastic term is of order $\epsilon$, and
dominates the one stored in the prestress which is of order
$\epsilon^2$. This is a crucial property, which is at the basis of
isostaticity and the criticality
of non-linear excitations in the compressed phase. 

Despite giving only a relatively small contribution, the contribution of the prestress term
is important. In fact, as usual in repulsive sphere systems, this is a
destabilizing term (it corresponds to a negative definite matrix) that,
though small, would imply unstable directions if the elastic part
is not full ranked. We conclude that the number of contacts
should be at least isostatic so that the total matrix is positive
definite and the minimum is stable.

\begin{figure}
\centering
\includegraphics[width=0.5\columnwidth]{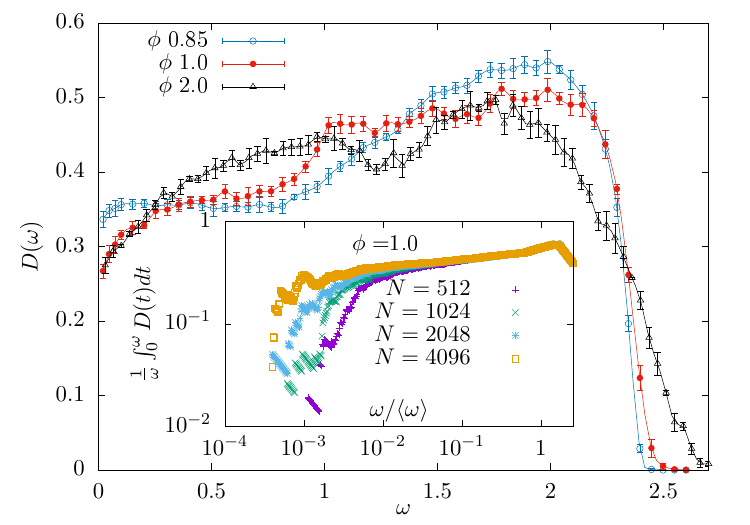}
\caption{Density of states (DOS) of the elastic part of the Hessian matrix of the regularized potential, see Eq.~(\ref{eq:2}), for different packing fraction above jamming in $d=2$ and with $N=4096$, averaged over $44$ samples for $\phi=0.85$, over $...$ samples for $\phi=1.0$, over $48$ samples for $\phi=2.0$. Inset: the finite size behavior of the left tail of the DOS is consistent with having a finite value for $D(\omega=0)$.}
\label{fig:evalues}
\end{figure}

The Hessian is therefore dominated by its isostatic random elastic part. Isostatic random matrices are gapless \cite{OSLN03,wyart2005geometric,yan2016variational,Pa14,FPUZ15,BPPS18} and characterized by an abundance of soft modes, their spectrum 
should  behave as $\lambda^{-1/2}$ at small
argument, where $\lambda$ represents the eigenvalues.
We measure the spectrum of the elastic term matrix, namely the spectrum of $\lim_{\epsilon \to 0} \epsilon {\cal H}_{ij}^{ab}$.  In Fig.\ref{fig:evalues}, we plot the corresponding density of states (DOS) with respect to the vibrational frequency $\omega=\sqrt{\lambda}$. Varying the density from $\phi=0.85$ to $\phi=2.0$, our numerical simulations are compatible with having a constant DOS for $\omega\rightarrow 0$.
In the appendix we develop a mean field theory for such behavior supporting this numerical finding.

\subsection{Non-linear excitations} \label{sec:excitations}
Further information can be gained considering 
a non-linear stability analysis for the
local minima of the PEL.  The data we
presented clearly shows that
minima are isostatic configurations where the distributions of both
positive and negative gaps display a power law behavior at small
argument.  At the same time the isostatic delta peak of marginally
satisfied gaps is accompanied by a contact force distribution which
has two pseudogaps close to forces equal to zero or one. The emergence
of these power laws controls the non-linear excitations that dominate the dynamics of the system 
when perturbing it away from such local minima.  
One can understand the nature of those excitations generalizing the
lines of reasoning employed in 
\cite{Wy12, LDW13} for the jamming point.  
The simplest excitations are the
ones in which isostaticity is off by one contact.  
There are here two possibilities, 
either separating two spheres in contact and opening a positive gap, or on the contrary pushing two spheres
in contact to make them overlap and create a negative
gap. The softest excitations are then the ones corresponding to
either very week contacts in the former case, or to 
contacts carrying a force close to one in the latter case. When such
contacts are removed, the system 
would become mechanically unstable unless a new contact forms in the
system and again we have two possibilities, either a gap closes, 
or an overlap relaxes to become a contact. Assuming that both processes occur with finite probability, we have  
$\th_+=\th_-$ and $\g_+=\g_-$. 
Following \cite{Wy12, LDW13}, 
one arrives at the scaling relation $\g_+=1/(2+\th_+)$ controlling the
critical exponents, which is verified 
by both our numerics and the mean field theory of \cite{FSU19}.

\section{Discussion and Conclusion}
{In this work we have described the emergence of a new critical phase
obtained when linear spheres are compressed above the jamming point.}
The criticality of local
minima of the PEL of linear soft spheres is described by a set of power laws controlling the
positive and negative gap distributions as well as contact forces.
The critical exponents controlling such distributions appear to be
numerically indistinguishable from the corresponding ones at jamming.
Furthermore, the critical behavior is again directly controlled by
isostaticity of local minima.  This is an interesting result that
opens the way to study jamming criticality in a different and
complementary way.  Indeed, typically, in order to look for the
critical properties of the jamming transition, one needs to fine-tune
the numerical simulations in order to be close to jamming.  Linear
soft spheres instead allow us to get to jamming-like critical
configurations just by looking at local energy minima which can be
obtained using standard numerical routines to minimize the energy. In
this case, no fine-tuning is needed. The rich physics
that we observe in linear spheres is due to isostaticity which
we robustly find with descent dynamics in local minima at finite $N$ \cite{HUZ19}.
{While the relevance of our work for materials as \emph{e.g.} soft colloids or granulars is left for future investigations,
the novelty of our results is directly manifested in the emergence of a new mechanism 
for marginal stability leading to criticality in a finite dimensional system.}

Our work opens a series of perspectives: on one hand,
it would be extremely interesting to characterize the rheology of
strained linear spheres \cite{BU16,RUYZ15}. A possible way to look for
that would be to perform similar experiments as in
Ref. \cite{CR00strain, FS17, SPB19} and to analyze the statistical
properties of avalanches.  On the other hand, it would be interesting
to investigate other concave penalty exponents $\alpha<1$, or more
complex potentials, to see if 
different non-linear criticality may arise.  Moreover, by switching on
temperature, one may investigate if marginal stability emerges at a
critical point, the Gardner transition \cite{KPUZ13,BCJPSZ16, BU18}.
Finally, further work is required to understand the 
behavior of bulk quantities such as energy and pressure close to
unjamming. Likely, this cannot be obtained from the scaling valid for
$\alpha>1$ \cite{durian1995foam,OSLN03,Wyart,GLS16}, and it may be important 
to study the leading corrections to this
scaling for $\alpha$ close to one. While our data hints at such phenomenology,
further investigations are needed.
A possible way to investigate this point would be to progressively compress
a configuration from jamming. The dynamics should be dominated by contacts carrying forces close to one becoming overlaps while small gaps becoming contacts with a net flux of gaps from the positive to the negative side of the distribution. How to describe such dynamics is left for future work.

% TODO: include funding information
\paragraph{Funding information}
This work was supported by ``Investissements d'Avenir" LabExPALM (ANR-10-LABX-0039-PALM) and by the Simons foundation 
(grants No.~454941, S.~Franz). SF is a member of the Institut Universitaire de France.

\begin{appendix}

\begin{figure}[h]
\centering
\includegraphics[scale=0.58]{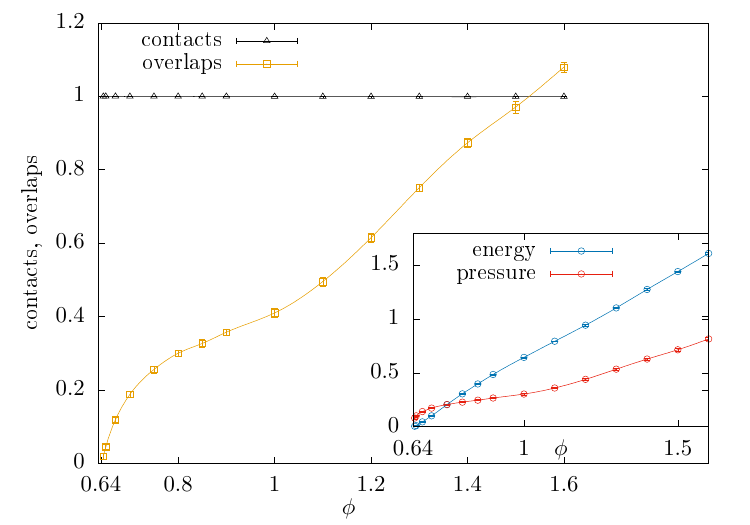}
\includegraphics[scale=0.58]{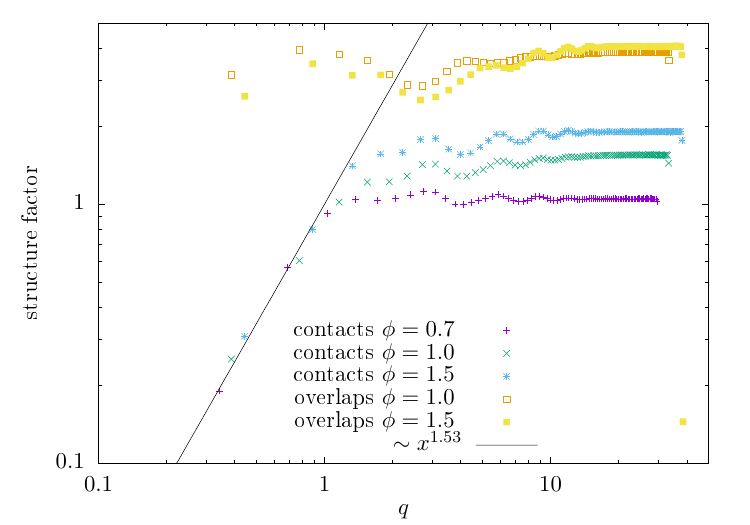}
\caption{\emph{Left panel}: isostaticity index and fraction of overlaps. While at all densities the minima are isostatic, the number of overlaps increases monotonically. In the inset, the corresponding energy and pressure.   \emph{Right panel}: structure factor of the local coordination of the contact and overlap network. While fluctuations of the local connectivity of the overlap network follow the central limit theorem, the ones of the contact network are hyperuniform.
Data produced with system size $N=1024$, dimensions $d=3$, averaged over $\sim 30$ samples for each density $\phi$.}
\label{fig:network3D}
\end{figure}

\begin{figure}[h]
\centering
\includegraphics[scale=0.5]{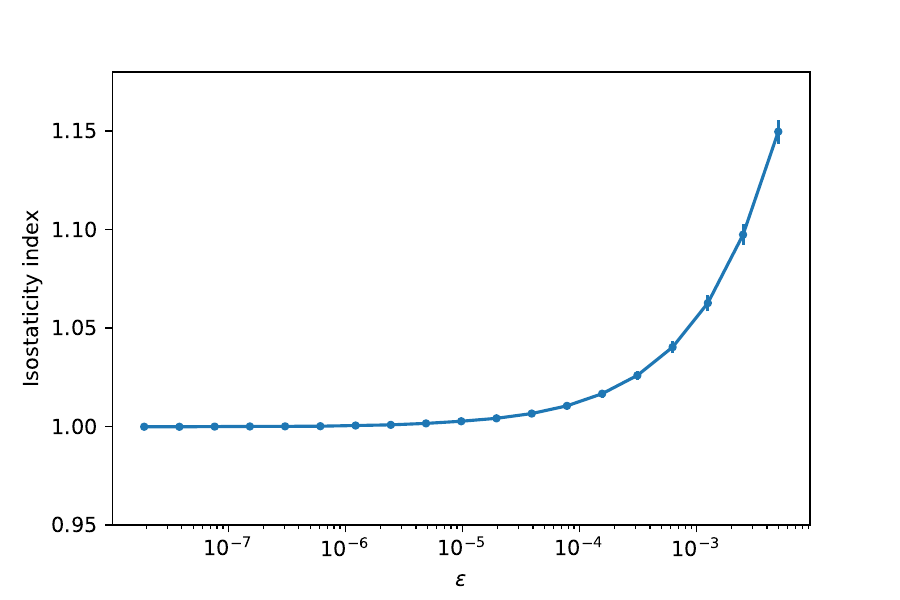}
\caption{The behavior of the isostaticity index as a function of the regularizer parameter $\epsilon$ for a system of $N=2048$ spheres in $d=2$ at $\phi=2$, averaged over $29$ samples. For $\epsilon\to 0$ the system sits in an isostatic minimum.}
\label{iso_eps}
\end{figure}

\begin{figure}[h]
\centering
\includegraphics[scale=0.55]{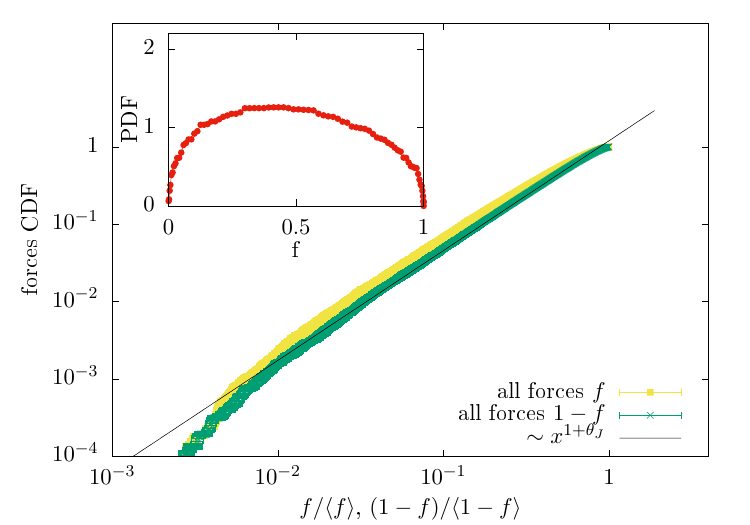}
\includegraphics[scale=0.55]{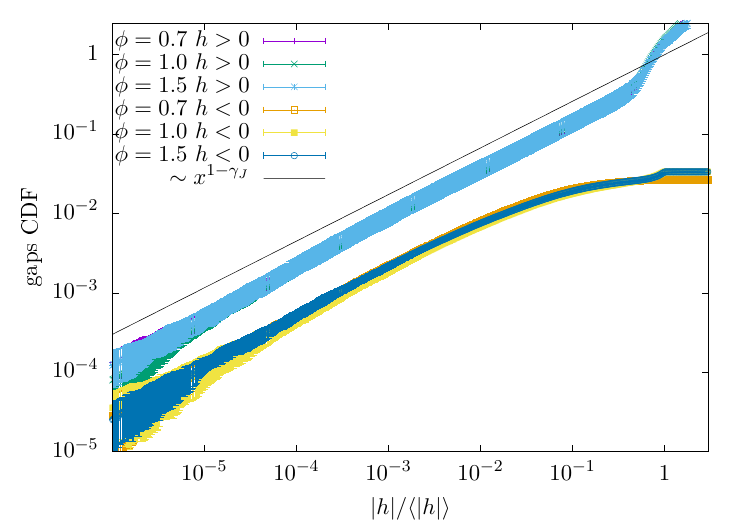}
\caption{\emph{Left panel}: Cumulative distribution of contact forces close to the two edges at $\phi=1.5$. The solid line corresponds to the mean field theory prediction. In the inset we plot the corresponding empirical distribution function. \emph{Right panel}: Cumulative distribution of small positive and negative gaps for different packing fractions. The solid line represents the mean field theory prediction.
Data produced with system size $N=1024$, dimensions $d=3$, averaged over $45$ samples for density $\phi=0.7$, $44$ samples for $\phi=1.5$, $27$ samples for $\phi=2.0$.}
\label{fig:gaps3D}
\end{figure}

\begin{figure}[h]
\centering
\includegraphics[scale=0.7]{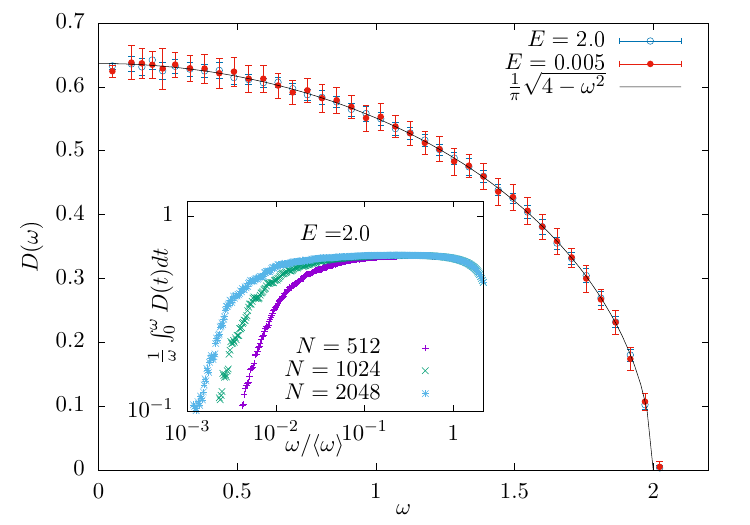}
\caption{Density of states of the linear perceptron optimization problem for two values of the energy at fixed $\alpha=5$ (at energy $E=2$ averaged over $30$ samples with system size $N=2048$, at energy $E=0.005$ averaged over $50$ samples with system size $N=1024$). The black line is the theoretical prediction given by Eq.~(\ref{Th_perc}). In the Inset we plot the left tail of the DOS for different sizes which shows that the behavior is compatible with having a positive DOS for $\omega \to 0$.}
\label{DOS_perc}
\end{figure}

\section{Properties of energy minima of linear spheres in three dimensions}
Here we report the results of numerical simulations of three dimensional linear soft spheres.
We consider $N=1024$ spheres with varying packing fraction $\phi$. With the polidispersity we are using, the jamming point is at $\phi\simeq 0.64$. We are interested in the properties of the PEL of linear spheres above this packing fraction. As for $d=2$, we find that our minimization algorithm produces isostatic minima, meaning  configurations in which an isostatic number of spheres are perfectly kissing. 
The properties of the contact network are the same as the ones for two dimensional packings.
In Fig.~\ref{fig:network3D}-left we plot the dependence of the fraction of contacts with respect to degrees of freedom (computed removing rattlers). In the same plot we also plot the fraction of overlaps. While we see that the system is isostatic at all packing fraction above jamming, the density of overlaps increases monotonically. In the inset we report the behavior of the energy and pressure as a function of $\phi$. 
In the right panel we plot also the structure factor of the local connectivity of the overlap and contact network. We show that while the overlap network obeys central limit theorem, the fluctuations of the number of contacts are suppressed and the structure factor goes to zero for small momenta.

Therefore both in $d=2$ and $d=3$ isostaticity is reached when minimizing the energy of the system. In order to see how this happens numerically, in Fig.~\ref{iso_eps} we plot the isostaticity index as a function of the regularization parameter $\epsilon$ (we plot data for $d=2$ for simplicity). We clearly see that as soon as the linear potential limit is reached, the system self organizes to sit on an isostatic minimum.

Therefore the main conclusion of this analysis is that, as for the jamming transition, the properties of soft spheres interacting with linear potential do not depend on the dimensionality of the system
(apart from local bucklers/rattlers effects).

Finally in Fig.\ref{fig:gaps3D} we plot the contact force and gap distribution.
Contact forces display two pseudogaps close to the edges $f=0,1$ while small gaps are controlled by a power law distribution. The critical exponents controlling these distributions appear to be the same (within numerical precision) with the ones of two dimensional linear spheres.

\section{Mean field theory of the density of states of the contact network matrix}
In the main text we have argued that the elastic part of the contact network matrix has a density of states $D(\omega)$ which goes to a positive constant for $\omega\to 0$ analogously to what happens at jamming. 
In this section we construct the mean field theory for such behavior.
We consider the spherical perceptron optimization problem with linear cost function studied recently in \cite{FSU19}, see also \cite{GG91,MEZ93, Gy01}.
This model is in the same universality class of linear soft spheres.
It is an optimization problem where a set of $N$ variables $x_i$ arranged in a vector $\underline x=\{x_1,\ldots, x_N\}$ lying on the sphere $|\underline x|^2=N$ are sought to minimize the cost function
\beq
H[\underline x] = \sum_{\mu=1}^{\alpha N} |h_\mu|\th(-h_\mu)
\eeq 
where the gaps $h_\mu$ are defined as
\beq
h_\mu=\frac 1{\sqrt{N}}\underline \xi^\mu\cdot \underline x -\sigma
\eeq
with $\underline \xi^\mu=\{\xi^\mu_1,\ldots \xi^\mu_N\}$ a set of $N$ dimensional vectors with components extracted from a Normal distribution and $\sigma$ a constant control parameter.
Local minima of the PEL are isostatic meaning that there is an isostatic number of gaps $h_\mu=0$ and characterized by critical power laws in the gap and forces distribution.
The Hessian of the non-analytic minima can be defined by smoothing out the singularity of the linear potential close to $h_\mu=0$ as we have done with linear soft spheres. Calling $\epsilon$ the smoothing parameter, the Hessian becomes
\beq
{\cal H}_{ij} = \frac 1{\epsilon N} \sum_{\mu: h_\mu=0} \xi^\mu_i\xi^\mu_j + \zeta\delta_{ij}
\eeq
being $\zeta$ a Lagrange multiplier needed to enforce the spherical constraint which plays here the same role of the prestress in spheres.
In the glassy phase, $\zeta<0$ and therefore for $\epsilon \to 0$ one needs to have isostatic minima.
Since the system is isostatic, assuming that the patterns are random and the only ingredient that matters for the statistics of the Hessian is the number of contacts \cite{FPUZ15}, we get that the spectrum of the Hessian is a given by a Marcenko-Pastur distribution for the eigenfrequencies $\omega$ given by
\beq
D(\omega) = \frac 1\pi \sqrt{4-\omega^2}\:.
\label{Th_perc}
\eeq
This result holds in the whole glassy phase regardless of the energy, being the glassy phase always isostatic.
In Fig.~\ref{DOS_perc} we plot the density of states as extracted from numerical simulation for $\alpha=5$ and at two different values of the energy and we compare it with the theoretical prediction.
\end{appendix}

%\bibliography{HS}

\begin{thebibliography}{10}
\providecommand{\url}[1]{\texttt{#1}}
\providecommand{\urlprefix}{URL }
\expandafter\ifx\csname urlstyle\endcsname\relax
  \providecommand{\doi}[1]{doi:\discretionary{}{}{}#1}\else
  \providecommand{\doi}{doi:\discretionary{}{}{}\begingroup
  \urlstyle{rm}\Url}\fi
\providecommand{\eprint}[2][]{\url{#2}}

\bibitem{LN98}
A.~J. Liu and S.~R. Nagel,
\newblock \emph{Jamming is not just cool any more},
\newblock Nature \textbf{396}(6706), 21 (1998),
\newblock \doi{https://doi.org/10.1038/23819}.

\bibitem{LN10}
A.~J. Liu and S.~R. Nagel,
\newblock \emph{The jamming transition and the marginally jammed solid},
\newblock Annu. Rev. Condens. Matter Phys. \textbf{1}(1), 347 (2010),
\newblock \doi{10.1146/annurev-conmatphys-070909-104045}.

\bibitem{OLLN02}
C.~S. O'Hern, S.~A. Langer, A.~J. Liu and S.~R. Nagel,
\newblock \emph{Random packings of frictionless particles},
\newblock Phys. Rev. Lett. \textbf{88}(7), 075507 (2002),
\newblock \doi{10.1103/PhysRevLett.88.075507}.

\bibitem{OSLN03}
C.~S. O'Hern, L.~E. Silbert, A.~J. Liu and S.~R. Nagel,
\newblock \emph{Jamming at zero temperature and zero applied stress: The
  epitome of disorder},
\newblock Phys. Rev. E \textbf{68}(1), 011306 (2003),
\newblock \doi{10.1103/PhysRevE.68.011306}.

\bibitem{CWBC98}
M.~E. Cates, J.~Wittmer, J.-P. Bouchaud and P.~Claudin,
\newblock \emph{Jamming, force chains, and fragile matter},
\newblock Phys. Rev. Lett. \textbf{81}(9), 1841 (1998),
\newblock \doi{10.1103/PhysRevLett.81.1841}.

\bibitem{TW99}
A.~V. Tkachenko and T.~A. Witten,
\newblock \emph{Stress propagation through frictionless granular material},
\newblock Phys. Rev. E \textbf{60}, 687 (1999),
\newblock \doi{10.1103/PhysRevE.60.687}.

\bibitem{LNSW10}
A.~J. Liu, S.~R. Nagel, W.~Van~Saarloos and M.~Wyart,
\newblock \emph{{The jamming scenario -- an introduction and outlook}},
\newblock In L.~Berthier, G.~Biroli, J.-P. Bouchaud, L.~Cipelletti and W.~van
  Saarloos, eds., \emph{Dynamical Heterogeneities and Glasses}. Oxford
  University Press,
\newblock \doi{10.1093/acprof:oso/9780199691470.003.0009} (2011).

\bibitem{GLS16}
C.~P. Goodrich, A.~J. Liu and J.~P. Sethna,
\newblock \emph{Scaling ansatz for the jamming transition},
\newblock Proc. Natl. Acad. Sci. U.S.A. \textbf{113}(35), 9745 (2016),
\newblock \doi{https://doi.org/10.1073/pnas.1601858113}.

\bibitem{LDW13}
E.~Lerner, G.~During and M.~Wyart,
\newblock \emph{Low-energy non-linear excitations in sphere packings},
\newblock Soft Matter \textbf{9}, 8252 (2013),
\newblock \doi{10.1039/C3SM50515D}.

\bibitem{CCPZ12}
P.~Charbonneau, E.~I. Corwin, G.~Parisi and F.~Zamponi,
\newblock \emph{Universal microstructure and mechanical stability of jammed
  packings},
\newblock Phys. Rev. Lett. \textbf{109}, 205501 (2012),
\newblock \doi{10.1103/PhysRevLett.109.205501}.

\bibitem{CCPZ15}
P.~Charbonneau, E.~I. Corwin, G.~Parisi and F.~Zamponi,
\newblock \emph{Jamming criticality revealed by removing localized buckling
  excitations},
\newblock Phys. Rev. Lett. \textbf{114}, 125504 (2015),
\newblock \doi{10.1103/PhysRevLett.114.125504}.

\bibitem{Wy12}
M.~Wyart,
\newblock \emph{Marginal stability constrains force and pair distributions at
  random close packing},
\newblock Phys. Rev. Lett. \textbf{109}, 125502 (2012),
\newblock \doi{10.1103/PhysRevLett.109.125502}.

\bibitem{CKPUZ14NatComm}
P.~Charbonneau, J.~Kurchan, G.~Parisi, P.~Urbani and F.~Zamponi,
\newblock \emph{Fractal free energies in structural glasses},
\newblock Nat. Commun. \textbf{5}, 3725 (2014),
\newblock \doi{10.1038/ncomms4725}.

\bibitem{CKPUZ14JSTAT}
P.~Charbonneau, J.~Kurchan, G.~Parisi, P.~Urbani and F.~Zamponi,
\newblock \emph{Exact theory of dense amorphous hard spheres in high dimension.
  iii. the full replica symmetry breaking solution},
\newblock JSTAT \textbf{2014}(10), P10009 (2014),
\newblock \doi{https://doi.org/10.1088/1742-5468/2014/10/P10009}.

\bibitem{CKPUZ17}
P.~Charbonneau, J.~Kurchan, G.~Parisi, P.~Urbani and F.~Zamponi,
\newblock \emph{Glass and jamming transitions: From exact results to
  finite-dimensional descriptions},
\newblock Annu. Rev. Condens. Matter Phys. \textbf{8}, 265 (2017),
\newblock \doi{10.1146/annurev-conmatphys-031016-025334}.

\bibitem{FP16}
S.~Franz and G.~Parisi,
\newblock \emph{The simplest model of jamming},
\newblock Journal of Physics A: Mathematical and Theoretical \textbf{49}(14),
  145001 (2016),
\newblock \doi{https://doi.org/10.1088/1751-8113/49/14/145001}.

\bibitem{FPUZ15}
S.~Franz, G.~Parisi, P.~Urbani and F.~Zamponi,
\newblock \emph{Universal spectrum of normal modes in low-temperature glasses},
\newblock Proc. Natl. Acad. Sci. U.S.A. \textbf{112}(47), 14539 (2015),
\newblock \doi{https://doi.org/10.1073/pnas.1511134112}.

\bibitem{FPSUZ17}
S.~Franz, G.~Parisi, M.~Sevelev, P.~Urbani and F.~Zamponi,
\newblock \emph{Universality of the sat-unsat (jamming) threshold in non-convex
  continuous constraint satisfaction problems},
\newblock SciPost Physics \textbf{2}(3), 019 (2017),
\newblock \doi{10.21468/SciPostPhys.2.3.019}.

\bibitem{FHU19}
S.~Franz, S.~Hwang and P.~Urbani,
\newblock \emph{Jamming in multilayer supervised learning models},
\newblock Phys. Rev. Lett. \textbf{123}(16), 160602 (2019),
\newblock \doi{10.1103/PhysRevLett.123.160602}.

\bibitem{HUZ19}
D.~Hexner, P.~Urbani and F.~Zamponi,
\newblock \emph{Can a large packing be assembled from smaller ones?},
\newblock Phys. Rev. Lett. \textbf{123}, 068003 (2019),
\newblock \doi{10.1103/PhysRevLett.123.068003}.

\bibitem{HLN17}
D.~Hexner, A.~J. Liu and S.~R. Nagel,
\newblock \emph{Two diverging length scales in the structure of jammed
  packings},
\newblock Phys. Rev. Lett. \textbf{121}(11), 115501 (2018),
\newblock \doi{10.1103/PhysRevLett.121.115501}.

\bibitem{FSU19}
S.~Franz, A.~Sclocchi and P.~Urbani,
\newblock \emph{Critical jammed phase of the linear perceptron},
\newblock Phys. Rev. Lett. \textbf{123}(11), 115702 (2019),
\newblock \doi{10.1103/PhysRevLett.123.115702}.

\bibitem{fire-BK06}
E.~Bitzek, P.~Koskinen, F.~G{\"a}hler, M.~Moseler and P.~Gumbsch,
\newblock \emph{Structural relaxation made simple},
\newblock Phys. Rev. Lett. \textbf{97}(17), 170201 (2006),
\newblock \doi{10.1103/PhysRevLett.97.170201}.

\bibitem{lbfgs-BN95}
R.~H. Byrd, P.~Lu, J.~Nocedal and C.~Zhu,
\newblock \emph{A limited memory algorithm for bound constrained optimization},
\newblock SIAM Journal on Scientific Computing \textbf{16}(5), 1190 (1995),
\newblock \doi{https://doi.org/10.1145/279232.279236}.

\bibitem{Wyart}
{Wyart, M.},
\newblock \emph{On the rigidity of amorphous solids},
\newblock Ann. Phys. Fr. \textbf{30}(3), 1 (2005),
\newblock \doi{10.1051/anphys:2006003}.

\bibitem{FSU20}
S.~Franz, A.~Sclocchi and P.~Urbani,
\newblock \emph{In preparation} .

\bibitem{PUZ20}
G.~Parisi, P.~Urbani and F.~Zamponi,
\newblock \emph{Theory of Simple Glasses: Exact Solutions in Infinite
  Dimensions},
\newblock Cambridge University Press,
\newblock \doi{10.1017/9781108120494} (2020).

\bibitem{wyart2005geometric}
M.~Wyart, S.~R. Nagel and T.~A. Witten,
\newblock \emph{Geometric origin of excess low-frequency vibrational modes in
  weakly connected amorphous solids},
\newblock EPL (Europhysics Letters) \textbf{72}(3), 486 (2005),
\newblock \doi{https://doi.org/10.1209/epl/i2005-10245-5}.

\bibitem{yan2016variational}
L.~Yan, E.~DeGiuli and M.~Wyart,
\newblock \emph{On variational arguments for vibrational modes near jamming},
\newblock EPL \textbf{114}(2), 26003 (2016),
\newblock \doi{https://doi.org/10.1209/0295-5075/114/26003}.

\bibitem{Pa14}
G.~Parisi,
\newblock \emph{Soft modes in jammed hard spheres (i): Mean field theory of the
  isostatic transition},
\newblock arXiv preprint  (2014),
\newblock \doi{arXiv:1401.4413}.

\bibitem{BPPS18}
F.~P. Benetti, G.~Parisi, F.~Pietracaprina and G.~Sicuro,
\newblock \emph{Mean-field model for the density of states of jammed soft
  spheres},
\newblock Phys. Rev. E \textbf{97}(6), 062157 (2018),
\newblock \doi{10.1103/PhysRevE.97.062157}.

\bibitem{BU16}
G.~Biroli and P.~Urbani,
\newblock \emph{Breakdown of elasticity in amorphous solids},
\newblock Nature physics \textbf{12}(12), 1130 (2016),
\newblock \doi{https://doi.org/10.1038/nphys3845}.

\bibitem{RUYZ15}
C.~Rainone, P.~Urbani, H.~Yoshino and F.~Zamponi,
\newblock \emph{Following the evolution of hard sphere glasses in infinite
  dimensions under external perturbations: Compression and shear strain},
\newblock Phys. Rev. Lett. \textbf{114}(1), 015701 (2015),
\newblock \doi{10.1103/PhysRevLett.114.015701}.

\bibitem{CR00strain}
G.~Combe and J.-N. Roux,
\newblock \emph{Strain versus stress in a model granular material: a devil's
  staircase},
\newblock Physical Review Letters \textbf{85}(17), 3628 (2000),
\newblock \doi{10.1103/PhysRevLett.85.3628}.

\bibitem{FS17}
S.~Franz and S.~Spigler,
\newblock \emph{Mean-field avalanches in jammed spheres},
\newblock Phys. Rev. E \textbf{95}(2), 022139 (2017),
\newblock \doi{10.1103/PhysRevE.95.022139}.

\bibitem{SPB19}
B.~Shang, P.~Guan and J.-L. Barrat,
\newblock \emph{Elastic avalanches reveal marginal behavior in amorphous
  solids},
\newblock Proc. Natl. Acad. Sci. U.S.A. \textbf{117}(1), 86 (2020),
\newblock \doi{10.1073/pnas.1915070117}.

\bibitem{KPUZ13}
J.~Kurchan, G.~Parisi, P.~Urbani and F.~Zamponi,
\newblock \emph{Exact theory of dense amorphous hard spheres in high dimension.
  {II}. {T}he high density regime and the gardner transition},
\newblock J. Phys. Chem. B \textbf{117}, 12979 (2013),
\newblock \doi{https://doi.org/10.1021/jp402235d}.

\bibitem{BCJPSZ16}
L.~Berthier, P.~Charbonneau, Y.~Jin, G.~Parisi, B.~Seoane and F.~Zamponi,
\newblock \emph{Growing timescales and lengthscales characterizing vibrations
  of amorphous solids},
\newblock Proc. Natl. Acad. Sci. U.S.A. \textbf{113}(30), 8397 (2016),
\newblock \doi{10.1073/pnas.1607730113}.

\bibitem{BU18}
G.~Biroli and P.~Urbani,
\newblock \emph{Liu-nagel phase diagrams in infinite dimension},
\newblock SciPost Physics \textbf{4}(4), 020 (2018),
\newblock \doi{10.21468/SciPostPhys.4.4.020}.

\bibitem{durian1995foam}
D.~J. Durian,
\newblock \emph{Foam mechanics at the bubble scale},
\newblock Phys. Rev. Lett. \textbf{75}(26), 4780 (1995),
\newblock \doi{10.1103/PhysRevLett.75.4780}.

\bibitem{GG91}
M.~Griniasty and H.~Gutfreund,
\newblock \emph{Learning and retrieval in attractor neural networks above
  saturation},
\newblock J. Phys. A \textbf{24}(3), 715 (1991),
\newblock \doi{https://doi.org/10.1088/0305-4470/24/3/030}.

\bibitem{MEZ93}
P.~Majer, A.~Engel and A.~Zippelius,
\newblock \emph{Perceptrons above saturation},
\newblock J. Phys. A \textbf{26}(24), 7405 (1993),
\newblock \doi{https://doi.org/10.1088/0305-4470/26/24/015}.

\bibitem{Gy01}
G.~Gy{\"o}rgyi,
\newblock \emph{Techniques of replica symmetry breaking and the storage problem
  of the mcculloch--pitts neuron},
\newblock Phys. Rep. \textbf{342}(4-5), 263 (2001),
\newblock \doi{https://doi.org/10.1016/S0370-1573(00)00073-9}.

\end{thebibliography}

\end{document}